\documentstyle[epsfig,emulateapj5]{aastex}

\received{}
\revised{}
\accepted{}

\shorttitle{X-ray Absorption}
\shortauthors{Fang et al.}

\reversemarginpar
\makeatletter

\newenvironment{figurehere}
   {\def\@captype{figure}}
   {}
\makeatother

\begin{document}

\newcommand{\cms}{$\rm cm^2$}

\title{High Resolution Spectroscopy of X-ray Quasars: Searching for
   the X-ray Absorption from the Warm-Hot Intergalactic Medium}

\author{Taotao Fang\altaffilmark{1}, Claude
   R.~Canizares\altaffilmark{2} and Herman
   L.~Marshall\altaffilmark{2}}
\altaffiltext{1}{Department of Astronomy, University of California,
   Berkeley, CA 94530, fangt@astro.berkeley.edu; {\sl Chandra} Fellow}
\altaffiltext{2}{Department of Physics and Center for Space
	Research, MIT,  77 Mass. Ave., Cambridge, MA 02139}

\begin{abstract}

We present a survey of six low to moderate redshift quasars with {\sl
  Chandra} and {\sl XMM}-Newton. The primary goal is to search for the
  narrow X-ray absorption lines produced by highly ionized metals in
  the Warm-Hot Intergalactic Medium. All the X-ray spectra can be
  fitted by a power law with neutral hydrogen absorption method. The
  residuals that may caused by additional emission mechanisms or
  calibration uncertainties are taken account by polynomial in order
  to search for narrow absorption features. No real absorption
  line is detected at above 3-$\sigma$ level in all the spectra. We discuss
  the implications of the lack of absorption lines for our
  understanding of the baryon content of the universe and metallicity
  of the intergalactic medium (IGM). We find that the non-detection of
  X-ray absorption lines indicates that the metal abundance of the IGM
  should be smaller than $ \sim 0.3$ solar abundance. We also discuss
  implications of the non-detection of any local ($z \sim 0$) X-ray
  absorption associated with the ISM, Galactic halo or local group,
  such as has been seen along several other lines of sight (LOS).  By
  comparing a pair of LOSs we estimate a lower limit on the hydrogen
  number density for the ($z \sim 0$) 3C~273 absorber of $n_H \gtrsim
  4\times 10^{-3}\,\rm cm^{-3}$.

\end{abstract}


\section{Introduction}

\begin{figure*}
\centerline{\psfig{file=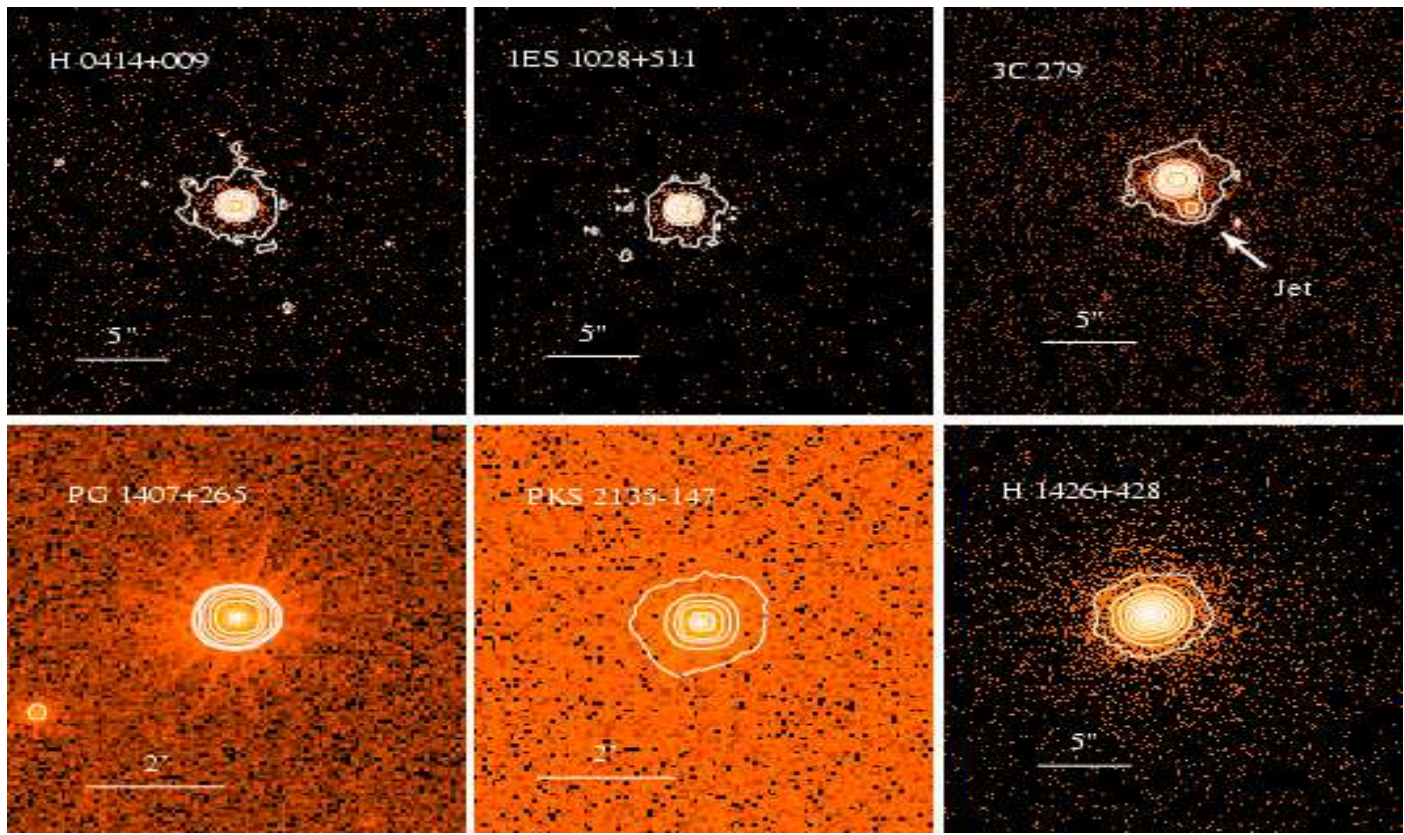,width=1\textwidth,height=0.45\textheight,angle=0}}
\caption[h]{X-ray images of all the targets, with north to the top
and east to the left. The white line in the
   bottom of each panel shows the angular size. For the 4 {\sl Chandra}
   targets, the zeroth order images are shown here; and EPIC images are
   shown for the two {\sl XMM}-Newton targets.}
\label{f1}
\end{figure*}

Cosmological hydrodynamic simulations predict that a significant
amount of baryons that reside in the Ly$\alpha$ forest at high
redshift were subsequently shock-heated to temperatures between
$10^5$--$10^7$ K, to form the so-called ``Warm-Hot Intergalactic 
Medium'', or WHIM
(see, e.g., \citealp{hgm98,cos99,dav01,fbc02,che03,vie03}). High
temperatures ionized most of the neutral hydrogen, making it impossible
to detect the WHIM gas as a low redshift Ly$\alpha$ forest. Thus, 
one of the central problems in studying the baryonic content of the
universe is to detect these ``missing baryons''.

In principal, such hot intergalactic gas can be
probed via both emission and absorption. However, the diffuse nature
of the hot gas makes it vary hard to directly image the WHIM gas (see,
e.g., \citealp{yos03,fan05}). On
the contrary, much effort has been put into studying the absorption
features produced by highly ionized metals in the spectra of
background quasars.

In the past few years, significant progress has been made in this
field, thanks to the advancement in high resolution spectroscopy
provided by the new generation of space telescopes. A significant
number of \ion{O}{6} absorption lines have been seen with the Far
Ultraviolet Spectroscopic Explorer ($\sl FUSE$) and the Hubble Space
Telescope ($\sl HST$) (see, e.g.,
\citealp{stl98,tsj00,tsa00,ssr02}). The distribution and derived
properties of these \ion{O}{6} lines are consistent with predictions
from simulations. In the X-ray band, \citet{fan02a},
\citet{mwc02}, \citet{mck03}, and \citet{nic05} reported on the detection of intervening \ion{O}{7}
and/or \ion{O}{8} absorption lines with {\sl Chandra} and {\sl
XMM}-Newton. \citet{nic02}, \citet{fsc03}, \citet{ras03},
\citet{cag04}, \citet{mck04}, and \citet{wil05} also reported the detection of $z \approx 0$ X-ray
absorption lines. These low redshift lines may be attributable, at
least in part, to the WHIM gas in our Local Group.

Starting from {\sl Chandra} and {\sl XMM}-Newton observation cycle 1,
we conducted a survey of X-ray bright quasars to systematically study
the potential X-ray absorption lines from the WHIM. In the first
round, we observed two high redshift quasars (PKS 2149-306 and
S5~0836+710 at $z \sim 2$, see \citealp{fan01}) and one low redshift
quasar (H~1821+643 at $z\sim 0.3$, see \citealp{fan02b}). In this
paper, we present the results from our second round of the survey,
which includes a total of six {\sl Chandra} and {\sl XMM} targets, and
a total exposure time of $\sim $ half-million seconds. These targets
were selected largely based on their X-ray flux levels, which were
determined by previous X-ray observations such as {\sl ROSAT} and {\sl
ASCA}. We also selected these targets because their low Galactic
absorption (except 1H~0414+009), and because their relatively simple
spectral shape (power law) without any intrinsic features. Two of
them were selected because their UV spectra showed \ion{O}{6}
absorption lines. In principle, we can either take a deep look at one
target, which can probe low column density absorbers, or observe
several lines-of-sight to increase the total pathlength. To maximize
the potential information we can gain, we selected the second approach.

This paper is organized as follows. We present the targets and
detailed data reduction procedures in section~\S2. We show the continuum
properties in section \S3. In section \S4 we discuss how we subtract
continuum and do narrow line analysis. Section \S5 is devoted to
detailed study of the results from narrow line analysis, and section 
\S~\ref{sec:sum} is summary and discussion.

\vbox{
\footnotesize
\begin{center}

\begin{tabular}{lcccc}
\multicolumn{5}{c}{~~~~~~~~Table 1: Target Parameters~~~~~~} \\
\hline \hline
Target & RA & Dec & Redshift & $N_{H}$\\
        & (J2000)    & (J2000) & & ($10^{20}$\cms)\\
\hline
PG~1407+265 & 14 09 23.9 & 26 18 21 & 0.940 & 1.38 \\
PKS~2135-147 & 21 37 45.2 & -14 32 55 & 0.200 & 4.77 \\
1H~0414+009 & 04 16 52.4 &  01 05 24 & 0.287 & 9.15 \\
1ES~1028+511 & 10 31 18.4 & 50 53 36 & 0.361 & 1.27 \\
3C~279 & 12 56 11.1 & -05 47 22 & 0.536  & 2.21\\
H~1426+428 & 14 28 32.6 & 42 40 21 & 0.129 & 1.36\\
\hline
\end{tabular}

\end{center}
}

\section{Observation and Data Reduction}

Our sample includes six quasars with redshifts ranging from $z \sim
0.13$ to $\sim$ 0.92. Four targets (3C~279, 1ES~1028+511, 1H~0414+009,
and H~1426+428) were observed with the High Energy Transmission
Grating Spectrometer (HETGS, see \citealp{can05}) \footnote{See
  http://space.mit.edu/HETG/.} onboard the {\sl Chandra} X-ray
telescope, and two (PG~1407+265 and PKS~2135-147) were observed with
the {\sl XMM}-Newton Observatory \footnote{See
  http://xmm.vilspa.esa.es/}. We list all the targets in Table~1 and
their relevant observational information in Table~2. These targets
were selected based on their strong X-ray flux obtained from previous
{\sl ROSAT} and/or {\sl ASCA} observations.

\vbox{
\footnotesize
\begin{center}

\begin{tabular}{lcccr}

\multicolumn{5}{c}{~~~~~~~~Table 2: Observation Log~~~~~~} \\
\hline \hline
Target & Telescope & Obs. ID & Obs. Date & Duration \\
             &                     &                &                   &  (sec) \\
\hline
PG~1407+265 & {\sl XMM} & 0092850501 & 2001-12-22 & 42,062 \\
PKS~2135-147 & {\sl XMM} & 0092850201 & 2001-04-29 & 59,850 \\
1H~0414+009 & HETGS & 2969  & 2002-08-01 & 52,000 \\
             &                     & 4284 & 2002-08-02 & 41,000 \\
1ES~1028+511 &  HETGS & 2970 & 2002-03-27 & 25,000 \\
              &                      & 3472 & 2002-03-28 & 75,000 \\
3C~279 &  HETGS & 2971 & 2002-03-21 & 107,000\\
H~1426+428 & HETGS & 3568 & 2003-09-08 & 102,000\\
\hline
\end{tabular}

\end{center}
}

\subsection{{\sl Chandra} Data Analysis}

\begin{figure*}
\centerline{\psfig{file=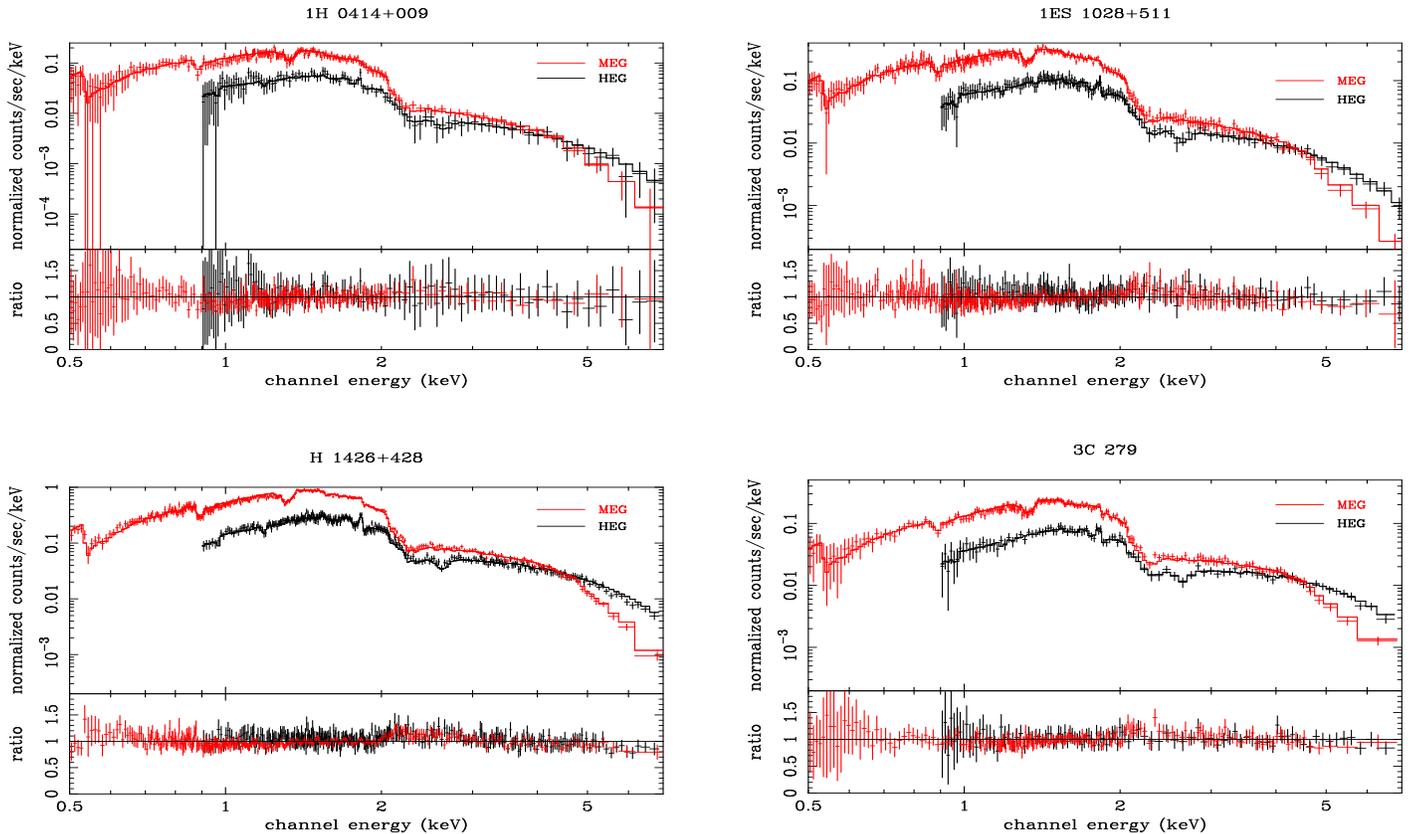,width=1\textwidth,height=0.45\textheight,angle=0}}
\caption[h]{{\sl Chandra} MEG (red line) and HEG (dark line)
   spectra. Solid lines in each plot represent the best fitted
   models. The bottom panel of each plot shows the ratio between data
   and model.}
\label{f2}
\end{figure*}

The {\sl Chandra} HETGS produces a zeroth order image at the aim-point
on the focal plane detector, the ACIS-S array, with higher order
spectra dispersed to either side (for ACIS-S, see \citealp{gar03}). 
For all four {\sl Chandra} observations, the telescope
pointing direction was offset $20\arcsec$ along $+$Y in order to move
the zeroth order off a detector node boundary, and the Science
Instrument Module (SIM) was moved toward the read-out row by about $3\
mm$ to get better ACIS energy resolution (for detailed instrument
setups, see the {\sl Chandra} Proposers' Observatory Guide, or
POG\footnote{See {\sl Chandra} Proposers' Observatory Guild (POG) at
http://asc.harvard.edu.}). Figure~\ref{f1}a, b, c, and d display the
zeroth-order images. We also label a size of $5\arcsec$ in each
panel. 1H~0414+009, 1ES~1028+511, and H~1426+428 appear to be point
sources with core sizes less than 2 -- 3 $\arcsec$; however, a large
scale jet is clearly presented in the zeroth order image of 3C~279
(south-east direction). Detailed study of this X-ray jet
\citep{mar03} will be presented in a separate paper.

{\sl Chandra} data were analyzed with the standard pipeline for the
{\sl Chandra} HETGS provided by the Chandra X-ray Center
(CXC)\footnote{See http://asc.harvard.edu/}. We use {\sl Chandra}
Interactive Analysis of Observations (CIAO) V3.0. The standard
screening criteria were applied to the data. We select photon events
with {\sl ASCA} grades 0, 2, 3, 4, 6 and excluded those with energies
above $10\ keV$. We also remove hot columns and bad pixels in each CCD
chip. and take into account the effect of ACIS absorption. The HETGS
consists of two different grating assemblies, the High Energy Grating
(HEG) and the Medium Energy Grating (MEG), and provides nearly
constant spectral resolution ($\Delta\lambda = 0.012 \AA$ for HEG and
$\Delta\lambda = 0.023 \AA$ for MEG) through the entire bandpass (HEG:
0.8-10 keV, MEG: 0.4-8 keV). The moderate energy resolution of the CCD
detector ACIS-S is used to separate the overlapping orders of the
dispersed spectrum. We added the plus and minus sides to obtain the
first order spectra of both grating assemblies. In cases where two
observations were performed for one target (1H~0414+009 and
1ES~1028+511), we use CIAO tool ``add\_grating\_spectra'' to add two
spectra together, and average the auxiliary response files (ARF).

\subsection{{\sl XMM}-Newton Data Analysis}

We obtained data for PG~1407+265 and PKS~2135-147 with all the
instruments onboard {\sl XMM}-Newton, and here we focus on EPIC
(including MOS1, MOS2 and PN) and RGS (including RGS1 and RGS2)
data. While with high spectral resolution RGS data will be analyzed to
study narrow line features, we also study EPIC data to obtain
information regarding the QSO continuum. Figure~\ref{f1}e and f
shows the MOS images of both QSOs.

\begin{figure*}
\psfig{file=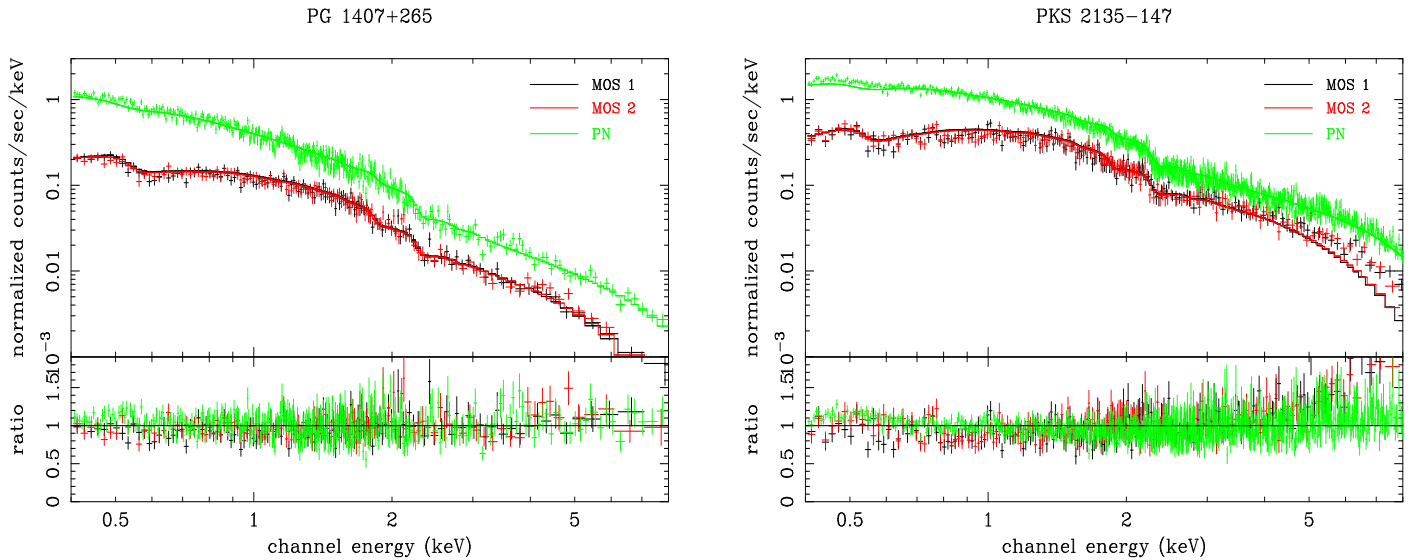,width=1\textwidth,height=0.3\textheight,angle=0}
\caption[h]{{\sl XMM}-Newton MOS1 (dark), MOS2 (red), and PN (green)
   spectra. Solid lines in each plot represent the best fitted
   models. The bottom panel of each plot shows the ratio between data
   and model.}
\label{f3}
\end{figure*}

All the EPIC observations were performed with the ``{\tt
PrimeFullWindow}'' and ``{\tt Imaging}'' data mode. A medium filter
was applied. Data were processed with the standard software, {\sl
XMM}-Newton Science Analysis System (SAS) V5.4 \footnote{see
http://xmm.vilspa.esa.es/}. We applied the standard SAS tasks ``{\tt
emchain}'' and ``{\tt epchain}'' to obtain event files for MOS and PN
data, respectively. We selected events with patterns between 0 and 12
for MOS and those with patterns between 0 and 4 for PN. All events
within central region with a radius of 30$\arcsec$ were extracted for
the point source spectra and areas between 30$\arcsec$ and 60$\arcsec$
for the background spectra. We also checked the pile-up level with
the SAS task ``{\tt epatploy}'' and found that there is no significant
pile-up between 0.4 and 10 keV, so we chose this energy range for the
subsequent spectral fitting. Finally, all the spectra were binned to
have a minimum of 20 counts per bin.

RGS 1 and 2 spectra were extracted using ``{\tt rgsproc}''. This SAS
task also provided a combined RMF and ARF file for each spectrum. To
achieve the highest signal-to-noise, we added both RGS (including the
first and second order spectra) together. The combined spectrum was
then grouped channel by channel to have a constant bin size of
$0.025\,\AA$, half of the width of the RGS FWHM, for the subsequent
narrow line analysis.


\begin{table*}[ht]
\begin{tabular*}{1.\textwidth}{@{\extracolsep{\fill}}lcccccc}
\multicolumn{6}{c}{~~~~~~~~Table 3: Spectral Fitting Parameters~~~~~~} \\
\hline \hline
Target & $N_{H}$          & Photon Index & $A_{pl}$$^a$ & 
$f_{2-10}$$^b$ & $L_{2-10}$$^c$ & $C-stat$/dof\\
        & ($10^{20}$\,\cms)  &   ($\Gamma$) & & & \\
\hline
3C~279 & 2.21$^d$  & $1.52\pm0.01$ & $3.00\pm0.03$ & 1.4 & 
$1.5\times10^{46}$ & 1125/950 \\
        & $5.0\pm1.0$ & $1.57\pm0.02$ & $3.19\pm0.07$ & 1.3 & 
$1.5\times10^{46}$ & 1104/949 \\
\hline
H~1426+428 & 1.36$^d$ & $1.80\pm0.01$ & $11.87\pm0.07$ & 3.6 & 
$1.7\times10^{45}$ &2098/950 \\
            & $9.7\pm0.5$ & $1.98\pm0.02$ & $14.35\pm0.20$ & 3.4 & 
$1.6\times10^{45}$ &1456/949 \\
\hline
1H~0414+009 & 9.15$^d$ & $2.49\pm0.02$ & $3.34\pm0.05$ & 0.4 & 
$1.2\times10^{45}$ &1079/950 \\
            & $11.88_{-1.32}^{+1.84}$ & $2.57\pm0.04$ &
$3.57\pm0.13$ &  0.4 & $1.2\times10^{45}$ & 1068/949 \\
\hline
1ES~1028+511 & 1.27$^d$ & $2.11\pm0.02$ & $4.83\pm0.05$ & 0.9 & 
$4.8\times10^{45}$ &1153/950 \\
            & $5.98_{-0.90}^{+0.70}$ & $2.24\pm0.02$ & $5.43\pm0.13$ &
0.9 & $4.7\times10^{45}$ & 1075/949 \\
\hline
PG~1407+265 & 1.38$^d$ & $2.24_{-0.02}^{+0.01}$ & $0.44\pm0.01$ & 0.08 &
$4.2\times10^{45}$ & 956/986 \\
\hline
PKS~2135-147 & 4.77$^d$ & $1.81\pm0.01$ & $1.65\pm0.01$ & 0.6 &
$6.2\times10^{44}$ & 2400/1521 \\
\hline
\label{tab:fit}
\end{tabular*}

\parbox{7in}{
\vspace{0.1in}
\indent
a. Flux at 1 keV (observer's frame) in units of 
$10^{-3}\rm\,photons\,cm^{-1}s^{-1}keV^{-1}$.\\
b. Flux between 2 --- 10 keV (observer's frame), in units of 
$10^{-11}\rm\,ergs\, cm^{-2}s^{-1}$.\\
c. Intrinsic luminosity between 2 --- 10 keV, in units of
$\rm\,ergs\,s^{-1}$.\\
d. $N_H$ fixed at the Galactic values.\\
}
\end{table*}

\section{Continuum Properties of Individual Source} \label{sec:cont}

Continuum can be fitted by a power law with absorption from neutral 
hydrogen. The flux $F(E)$ can be expressed as
\begin{equation}
F(E) = A_{pl}\left(\frac{E}{1\rm\,keV}\right)^{-\Gamma}\exp 
\left[-\sigma(E)N_{H}\right].
\label{eq:flux}
\end{equation}
Here $A_{pl}$ is the normalization at 1 keV, $\Gamma$ is the photon
index, $E$ is the photon energy, $\sigma(E)$ is the photoionization
cross section, and $N_H$ is the neutral hydrogen column
density. Table~3 lists the fitting parameters for all the targets and
inferred fluxes and luminosities. Fitting was performed with XSPEC
V11.3 \footnote{see
http://heasarc.gsfc.nasa.gov/docs/xanadu/xspec/}. For each {\sl
Chandra} data set we fit with with two models: (a) a power law with a
fixed neutral hydrogen absorption at the Galactic level, and (b) a
power law with a variable hydrogen absorption, and we fit both HEG and
MEG data simultaneously (see Figure~\ref{f2}). The fitting is
performed between 0.9 and 8 KeV HEG, and 0.5 and 8 keV for MEG. In the
top four targets in Table~3, the first rows show the fitted results
from model (a) and the second rows show results from model (b). For
the two {\sl XMM} targets, since we cannot constrain the hydrogen
column density ($N_H$), we fix them at the Galactic value and fit the
data with model (a) only. The fitting performed with all the three
instruments (MOS1, MOS2 and PN) simultaneously (see Figure~\ref{f3}),
between 0.4 and 10 Kev. We emphasize that while from the power law
fit some spectra show residuals that suggest additional emission
mechanisms, we are not trying to understand the emission processes
but only need to fit continuum to search for absorption lines. In
the next section, these residuals will be taken into account by
polynomials in order to search for narrow features. Since in many
places of the spectra we have to deal with small number of photons
per bin, we fit the spectrum using $C$-statistics in stead of the
conventional $\chi^2$-statistics \citep{cas76}. In Table~3 and
following sections, errors are quoted as 90\% confidence unless
otherwise mentioned. 

Comments on individual sources:
\begin{itemize}
\item{\it 3C~279}: This source is a radio loud quasar (RLQ) and was
identified as the first superluminal source \citep{whi71}. Our
{\sl Chandra} observation, for the first time, revealed the jet in
X-ray band (for detail see \citealp{mar03}). While both {\sl ROSAT}
(\citealp{cfg97,sam97}) and {\sl ASCA} \citep{rtu00}
observations showed similar photon indices ($\Gamma \sim 1.8$), we
obtained a relatively flat spectrum with {\sl Chandra} ($\Gamma \sim
1.5 - 1.6$). The fluxes and luminosities are consistent among these
observations.

\item{\it H~1426+428}: This is a BL Lac object and was extensively
   observed with {\sl HEAO-1} \citep{woo84}, {\sl EXOSAT}
   \citep{rem89}, {\sl BBXRT} \citep{mad92}, {\sl
   ROSAT} and {\sl ASCA} (see,e.g.,\citealp{sam97a}), and {\sl
   BeppoSAX} \citep{cos01}. Our result
   shows that the fitted $N_H$ is significantly higher than the
   Galactic value, which is consistent with results from {\sl BBXRT}
   and {\sl ASCA} but not with {\sl ROSAT} \citep{sam97a}. The
   observed flux between 2 --- 10 keV varied from $\sim 2$ to $\sim
   5 \times 10^{-11}\rm\,ergs\, cm^{-2}s^{-1}$ among various
   observations, while the photon index varies slightly around 2. While
   {\sl BBXRT} reported an X-ray absorption feature at $\sim 0.66$ keV
   (\citealp{mad92,sam97a}), we could not confirm
   this in our {\sl Chandra} MEG spectrum, consistent with results from
   recent {\sl XMM}-Newton observations \citep{blu04}.

\item{\it H~0414+009}: This target is a BL Lac object associated with
   a galaxy cluster of Abell richness of 0 \citep{fpt93}. {\sl
   BeppoSAX} and {\sl ROSAT} observations \citep{wol98} showed similar
   photon indices. However, we obtain a lower 2 --- 10 keV flux,
   indicating flux variation by a factor of $\sim 2$. We also find a
   slightly high $N_H$, compared with the Galactic value.

\item{\it 1ES~1028+511}: The redshift of this BL Lac object has been
   accurately determined as $z=0.361$, based on the measurement of two
   \ion{Ca}{2} absorption lines \citep{pol97}. {\sl ROSAT}
   observation showed a power law spectrum with $\Gamma = 2.43\pm0.20$
   and an unabsorbed flux of $6.16\times10^{-11}\rm\,ergs\,
   cm^{-2}s^{-1}$ between 0.5 --- 2.4 keV. This source was also among
   sources that listed in the {\sl ASCA} Medium Sensitive Survey
   \citep{ued01} with a flux of $7.77\times10^{-12}\rm\,ergs\,
   cm^{-2}s^{-1}$ between 2 --- 10 keV, assuming a photon index of 1.7.

\item{\it PG~1407+265} It's one of the brightest quasars at redshift around
   1. There is weak evidence for Damped Ly$\alpha$ Absorption and for
   Lyman Limit System in its optical
   continuum~\citep{lwt95}. A {\sl HST} observation revealed intrinsic
   narrow absorption features of \ion{O}{6} and Ly$\alpha$ in the
   quasar rest frame \citep{gan01}. Previous {\sl ASCA}
   observation showed a spectrum with similar photon index ($\Gamma
   \sim 2$) but stronger 2 --- 10 keV flux \citep{rtu00}.

\item{\it PKS~2135-147}: This is a typical double-lobe, radio source
   (\citealp{mha78,ghu84}). X-ray
   observations with {\sl Einstein} \citep{wel87} and {\sl
   EXOSAT} \citep{srv91} reported a photon
   index consistent with our observation; however, {\sl ROSAT}
   observations
   \citep{rmb96} showed a much steeper spectrum, with
   $\Gamma \sim 2.5$. We also obtained a lower X-ray flux, compared
   with previous observations. Optical and
   UV spectra show strong absorption lines of \ion{O}{6}, \ion{N}{5}
   and Ly$\alpha$ at $z_{abs} \sim z_{em}$~\citep{ham97}. These
   absorption lines also lie
   near the center of a small cluster and is very close to three
   galaxies inside that cluster. It is still unclear whether these
   absorption lines are intrinsic or due to the intervening systems.

\end{itemize}

\section{Narrow Line Analysis} \label{sec:narrow}

Cosmological simulations predict that typical X-ray absorption lines
from the WHIM gas have line widths on the order of m\AA~due to
velocity broadening (see, e.g.,
\citealp{hgm98,fbc02,che03}). On the observation side, the detection
of such narrow features relies on a careful measurement of the
continuum level. Using 1ES~1028+511 as an example, in
Figure~\ref{f4} we show how we obtain a continuum-subtracted
spectrum.

Figure~\ref{f4} shows a small portion of the raw count spectrum of
1ES~1028+511 plotted against wavelength. The solid dark line in the
top panel shows the MEG first order counts between 5 and 7 \AA\,, with
a bin size of 0.02\AA\,. The bin size is chosen in such a way that
there are {\it at least} two bins across the FWHM of the point spread
function (PSF) of the  instrument. In this way the bin size is
0.005\AA\, for {\sl Chandra} HEG, 0.01\AA\, for MEG and 0.025\AA\, for
{\sl XMM} RGS \footnote{See
http://asc.harvard.edu/proposer/POG/html/HETG.html and
http://xmm.vilspa.esa.es/external/xmm\_user\_support/documentation/uhb/node45.html
for the width of instrument PSF of {\sl Chandra} HETG and {\sl XMM}
RGS, respectively.}. The red line (labeled model 1) shows the
continuum from a power law plus neutral hydrogen absorption, with the
fitting parameters adopted from Table~\ref{tab:fit}. The bottom panel
gives $\chi$, which effectively is the corresponding Gaussian sigma of
the Poisson distribution \citep{geh86}. 

\begin{figure*}
\psfig{file=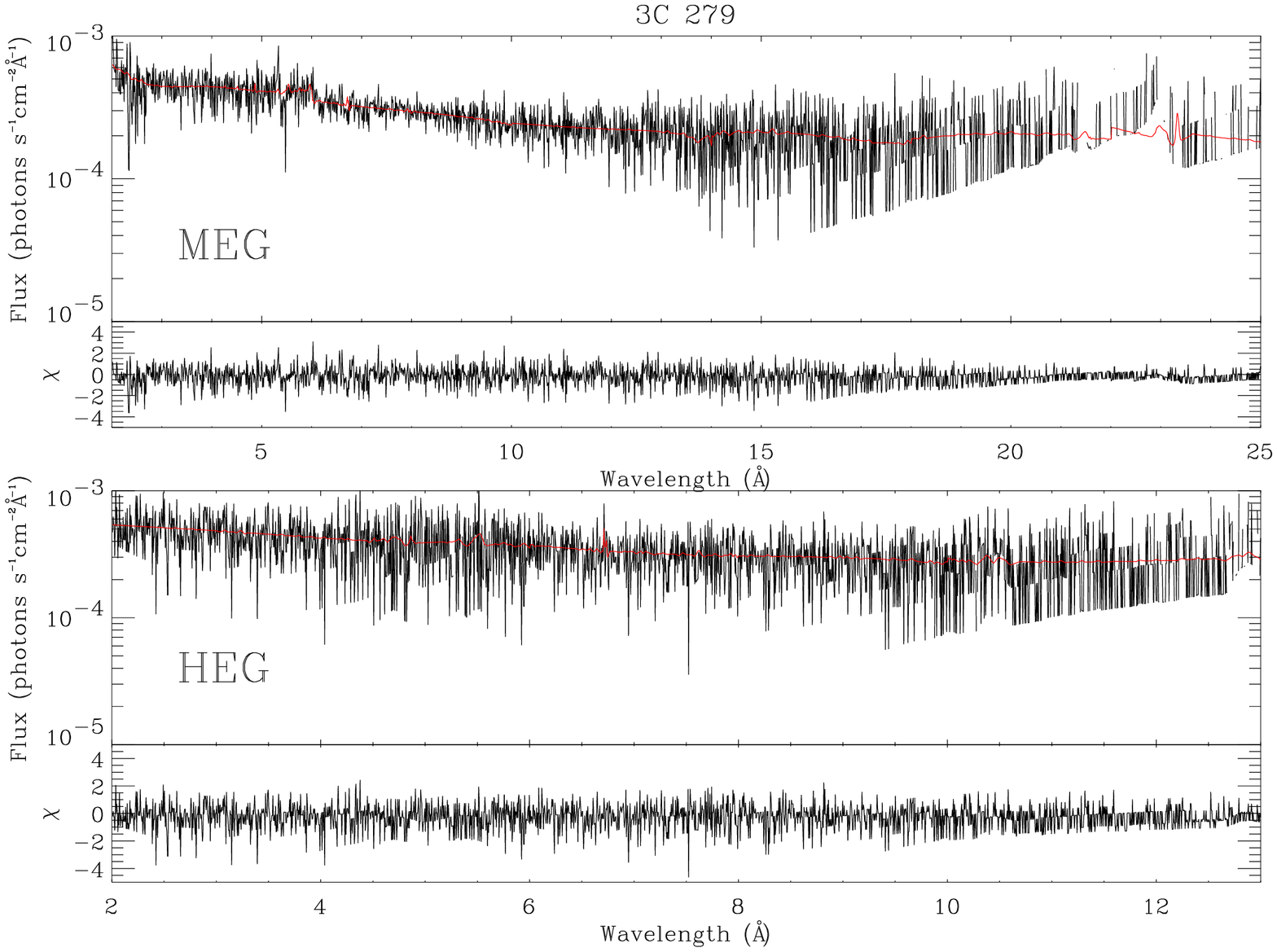,width=0.45\textwidth,height=0.75\textheight,angle=90}
\caption[h]{{\sl Chandra} MEG (top panel) and HEG (bottom panel) spectra of
   3C~279. Red lines show best fitted spectra from model 2. The bottom
   of each panel show the $\chi$.}
\label{f5}
\end{figure*}

\begin{figure*}
\psfig{file=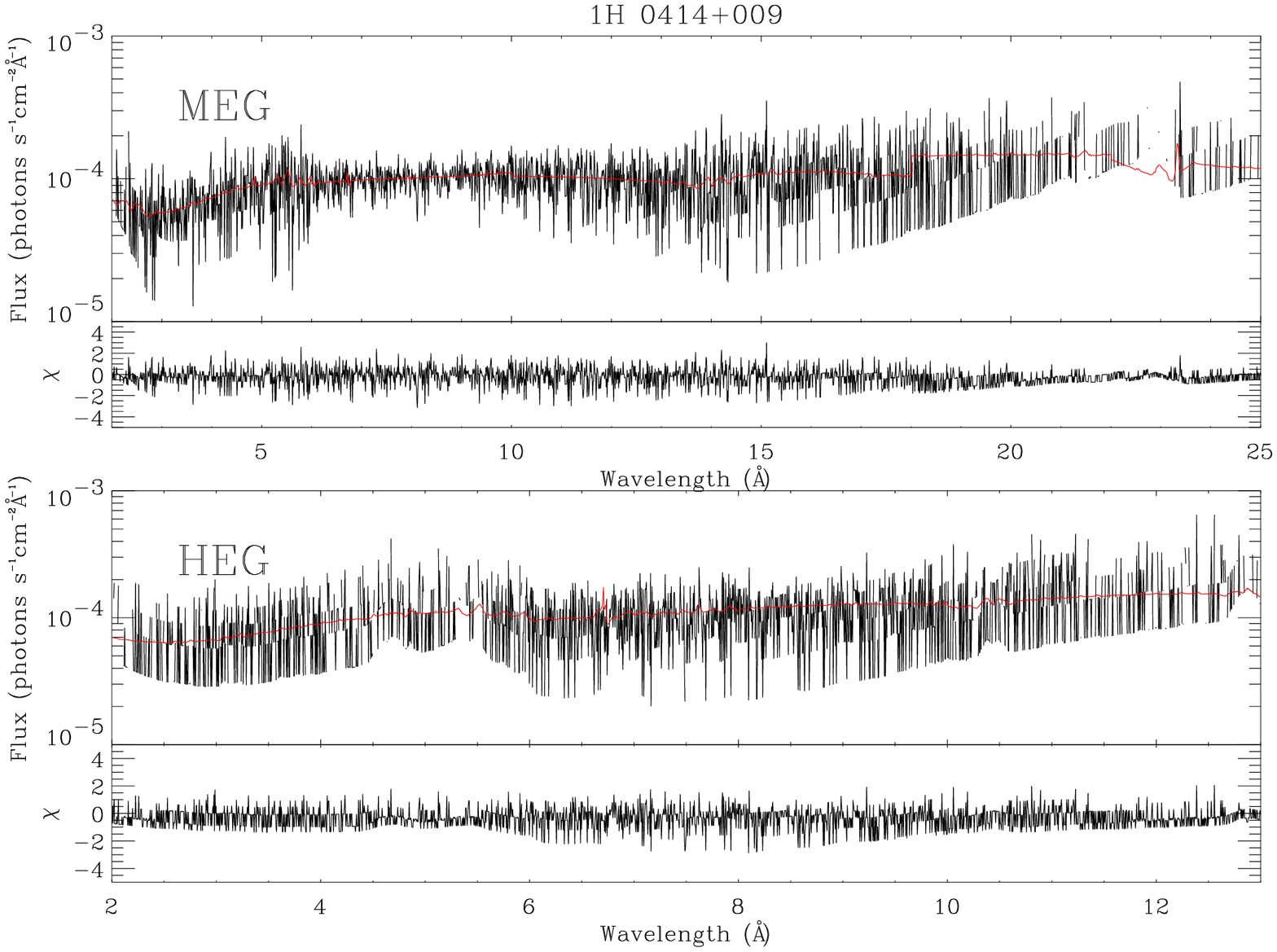,width=0.45\textwidth,height=0.75\textheight,angle=90}
\caption[h]{Same as Figure~\ref{f5} but for 1H~0414+009.}
\label{f6}
\end{figure*}

\begin{figure*}
\psfig{file=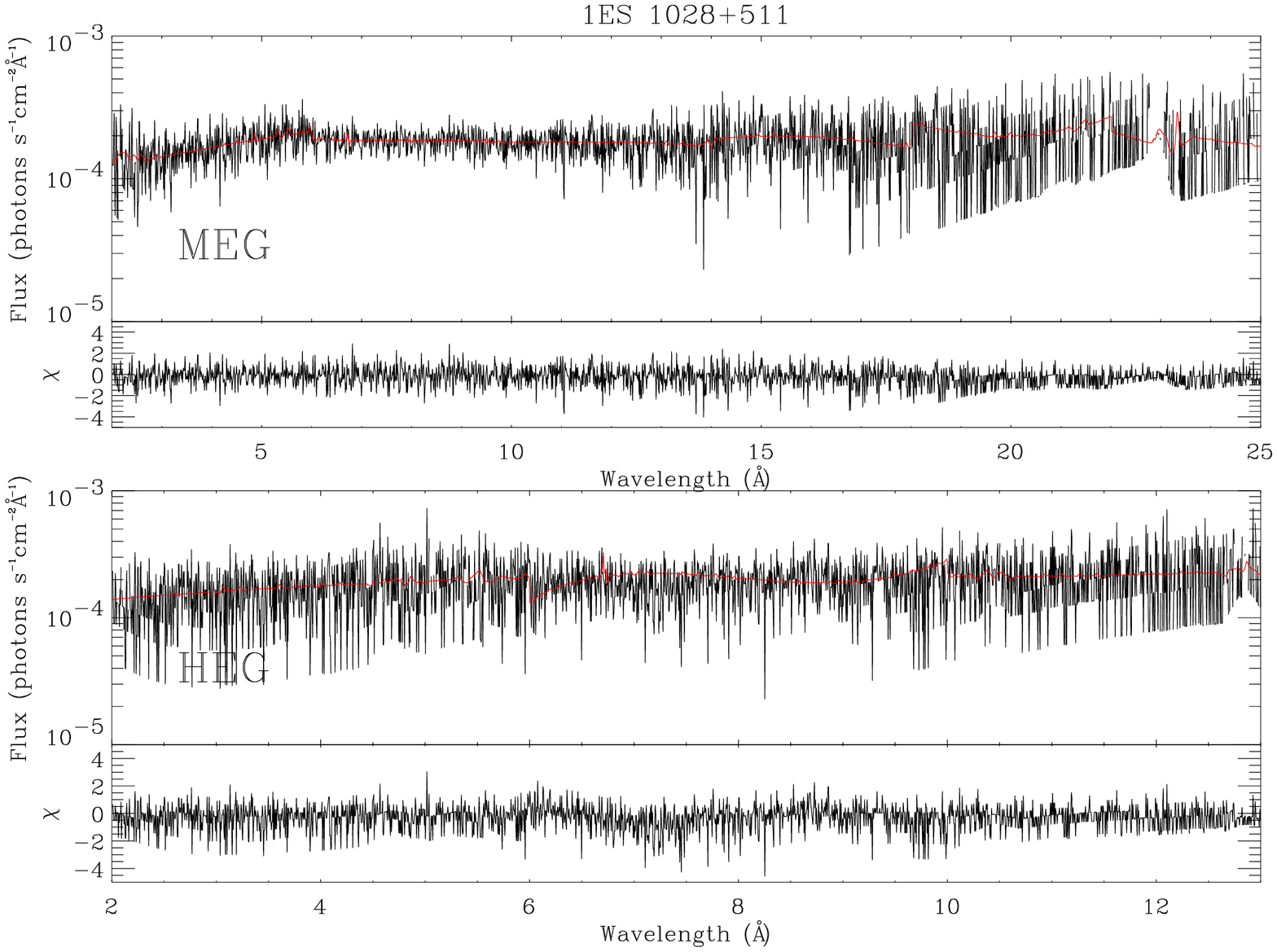,width=0.45\textwidth,height=0.75\textheight,angle=90}
\caption[h]{Same as Figure~\ref{f5} but for 1ES~1028+511.}
\label{f7}
\end{figure*}

\begin{figure*}
\psfig{file=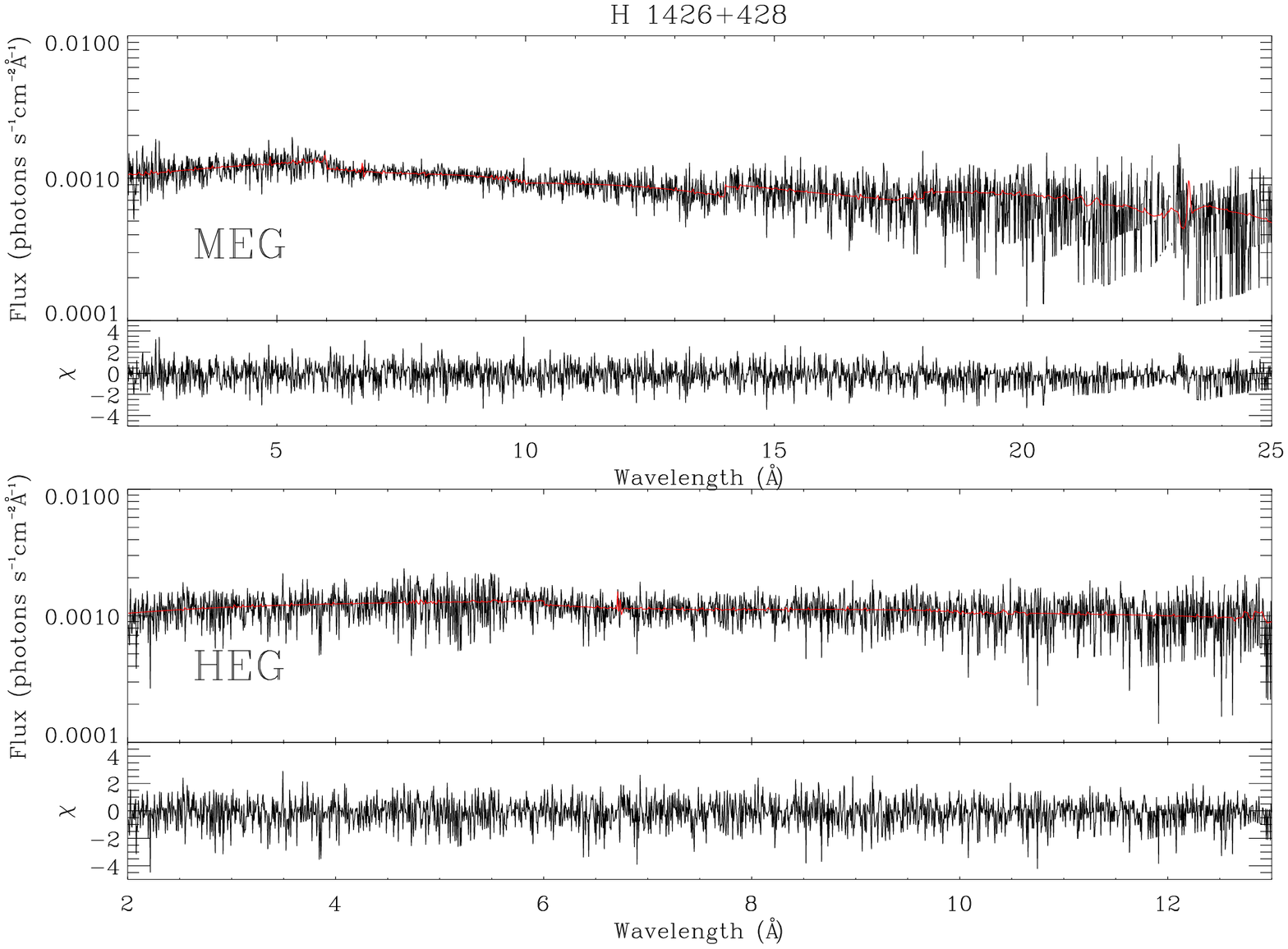,width=0.45\textwidth,height=0.75\textheight,angle=90}
\caption[h]{Same as Figure~\ref{f5} but for H~1426+428.}
\label{f8}
\end{figure*}

\begin{figure*}
\psfig{file=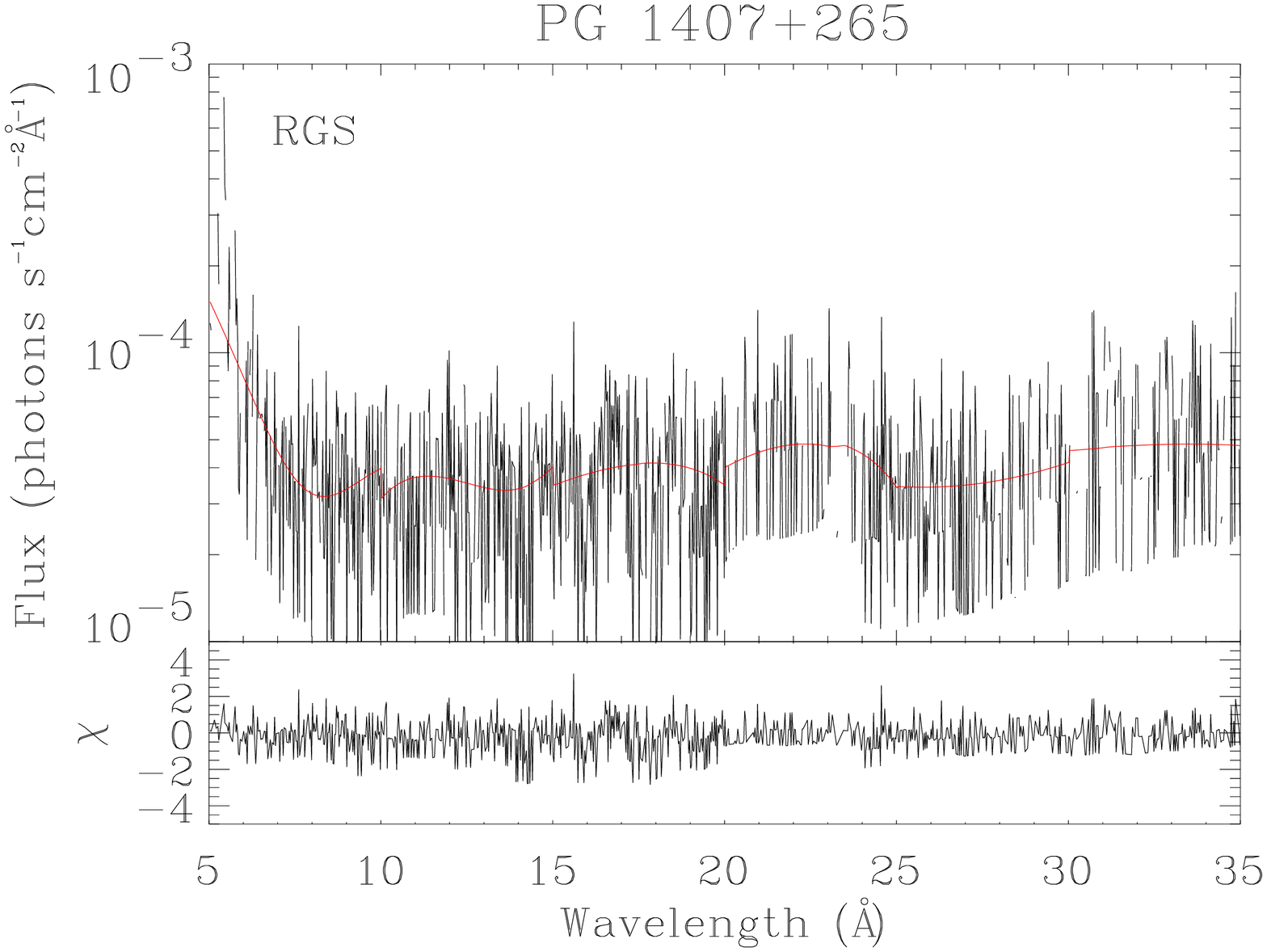,width=0.45\textwidth,height=0.75\textheight,angle=90}
\caption[h]{{\sl XMM}-Newton RGS spectrum of PG~1407+265. Lines are
   the same as those in Figure~\ref{f5}.}
\label{f9}
\end{figure*}

\begin{figure*}
\psfig{file=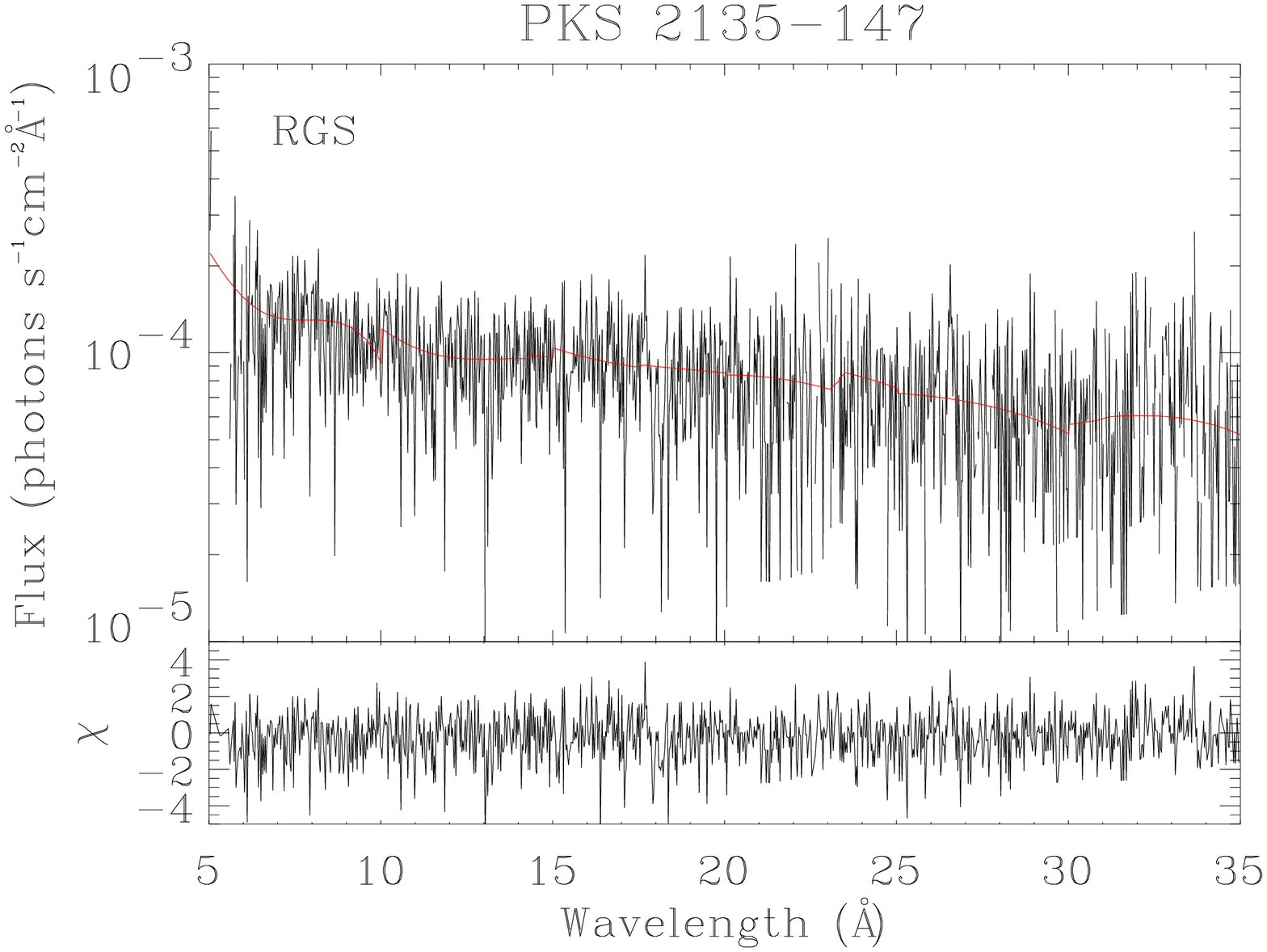,width=0.45\textwidth,height=0.75\textheight,angle=90}
\caption[h]{{\sl XMM}-Newton RGS spectrum of PKS~2135-147. Lines are
   the same as those in Figure~\ref{f5}.}
\label{f10}
\end{figure*}

While on scales larger than $\sim$ 1 \AA\, the absorbed power law provides a good fit to the overall
spectrum, in some local regions it will either over or under-estimate
the observed counts due to instrumental or intrinsic fluctuations. Clearly, in the bottom panel of Figure~\ref{f4}, there
are more red bins above zero than bins below zero, which means in the
5 -- 7\AA~region, model 1 underestimates the observed counts. The discrepancy
amounts to $\sim$ 10\%. Using such a model could compromise the
search for narrow absorption or emission lines. To eliminate this problem,
we first obtain the residual spectra by subtracting model
1. We then divide the whole residual spectra into several small
regions, and in each region we fit the residual with a 3-order
polynomial. To avoid significant deviation caused by one or two
randomly high or low $\chi$ bins, we ignore all the bins with $|\chi|
> 3$ before fitting. The selected spectral ranges are 2 and 30
\AA\,for MEG, 2 and 14 \AA\,for HEG, and 5 and 35 \AA\,for RGS. The
region size is 4 \AA,\,4 \AA,\,and 5 \AA,\,for MEG, HEG, and RGS,
respectively. This method allows us to minimize the fluctuations at
large scales, and still preserve features with width narrower than 1
to 2 \AA,\, which approximately is the region size divided by the
order of the fitted polynomial. We find the histogram distribution of
the $\chi$ obtained in this method can be well fitted by a Gaussian
distribution. We call this method ``model 2'' and plot it in blue in
Figure~\ref{f4}. Clearly, model 2 provides a significantly better
fit to the observed continuum. From Figure~\ref{f5} to ~\ref{f10} we showed the data and the
best fitted results from ``model 2''.

\vskip0.5cm
\begin{figurehere}
\psfig{file=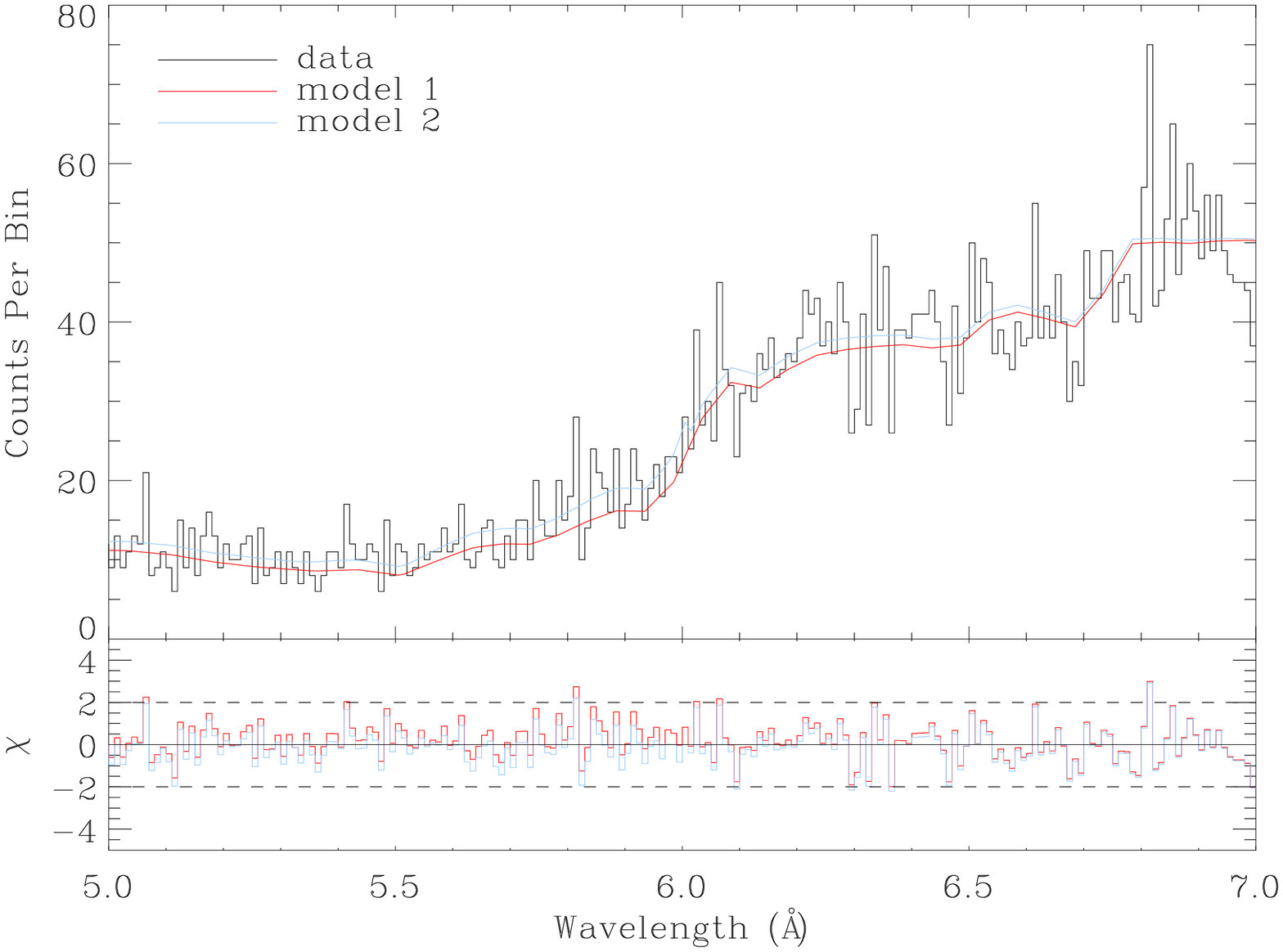,width=0.35\textwidth,height=0.33\textheight,angle=90}
\caption[h]{A sample of data (dark) and comparison between two models.
Model 1 (shown in red) adopted simplily the best fitted model from
Table~\ref{tab:fit}, and model 2 (blue) uses an local fitting
technique. Bottom panel shows the $\chi$, the equivalent Gaussian
$\sigma$ of Poisson distribution of each model. Clearly, model
2 is a better fit to data.}
\label{f4}
\end{figurehere}
\vskip0.5cm

The next step is to identify any potential absorption (or emission)
features in the spectrum. Our search criterion is still based on
$\chi$: we search for any feature which has at least two
continuous bins with $|\chi| > \chi_{min}$. Assuming the 
signal-to-noise ratio (SNR) in the first bin is
$\chi_1$ and in the next bin is $\chi_2$, Total SNR in these two
bins follows $\chi = (\chi_1+\chi_2)/\sqrt{2}$. To ensure a minimum
detection of at least 3$\sigma$, we need to have both $\chi_1$ and
$\chi_2 \lesssim -2$ for absorption features and both $\chi_1$ and
$\chi_2 \gtrsim 2$ for absorption features. We
label features identified with this requirement as potential
absorption or emission lines. We then fit the continuum-subtracted
residual with a Gaussian line profile around the features we
identified. The subtraction and fitting are performed with the
software package ISIS (Interactive Spectral Interpretation System, see
\citealp{hde00}) \footnote{see http://space.mit.edu/ASC/ISIS/}

Combining all the data together, we have a total of $\sim$ 20,000
bins. Based on the above criteria, we find a total of 28 possible
absorption features. Other stringent criteria are enforced to
eliminate false detections, including that (1) a true feature should
appear in both MEG and HEG spectra for {\sl Chandra} targets; (2) if a
feature appears in MEG only (at wavelength longer than 14 \AA), it
should appear in both $+1$ and $-1$ orders; (3) for RGS, a true
detection should appear in both RGS-1 and RGS-2; and (4) we should
avoid regions right at the instrument features - these regions
typically show large discrepancies between model and data. With these
criteria no real feature is detected. We also searched for emission
features and find no significant detections.

Other methods have been applied to fit the continuum and detect narrow
absorption features. For example, \citet{mck04} added an inverted
Gaussian to a physically-motivated continuum model and searched for
absorption features by monitoring changes in fit statistics. To
compare these two methods, we make a faked spectrum by adding narrow
features to an absorbed power law continuum and then apply both
methods. We find that both methods can effectively identify narrow
absorption lines with nearly the same significance. However, our
method is more robust when the underlying continuum is more complex
and has unknown components, and when there are uncertainties in
continuum calibration.

\section{Discussion} \label{sec:dis}

\subsection{Intervening IGM Absorption}

Based on the non-detection of any significant absorption features, we 
can determine
the minimum detectable ion column densities at a certain SNR
level. Given the quasar continuum flux $f_X$, the minimum detectable
line equivalent width ($W_{\lambda}$) is
\begin{equation}
W_{\lambda} = SNR \times \left(\frac{\lambda}{f_XRAT}\right)^{\frac{1}{2}},
\end{equation}
where $\lambda$ is wavelength, $R\equiv \lambda/\Delta \lambda$ is 
the resolving power,
$A$ is the effective area and $T$ is the exposure time. We expect
weak, unresolved features from the linear part of the 
curve-of-growth. The minimum detectable
column density ($N_i$) and equivalent width $W_{\lambda}$ then follow the
linear relationship \citep{spi78}
\begin{equation}
\frac{W_{\lambda}}{\lambda} = 8.51 \times 10^{-13} N_i \lambda f.
\end{equation}
Here $f$ is the oscillator strength.

In Table~4 we list the minimum detectable column densities of
\ion{O}{7} and \ion{O}{8} along the line-of-sight towards these six
targets. We adopt a SNR of 3 here. The rest frame wavelengths of
\ion{O}{7} and \ion{O}{8} are 21.6019\,\AA\, and 18.9689\,\AA,
respectively \citep{ver96}. Since {\sl Chandra} HEG has
no effective area in the longer wavelength of both transitions, we use
MEG only for the four {\sl Chandra} targets. Since MEG has no
effective area above 25 -- 26\,\AA, we can probe the line detection
up to $z \sim min(z_i,0.35)$ for \ion{O}{8} and $z \sim min(z_i,0.2)$ for
\ion{O}{7}, where $z_i$ is the redshift of individual sources.

\vbox{
\footnotesize
\begin{center}

\begin{tabular}{lcc}
\multicolumn{3}{c}{~~~~~~~~Table 4: 3$\sigma$ Limits on Column Density$^a$~~~~~~} \\
\hline \hline
Target & \ion{O}{8} & \ion{O}{7} \\
\hline
1H~0414+009 & 1.87 & 1.12 \\
1ES~1028+511 & 1.44 & 0.86 \\
H~1426+428 & 1.00 & 0.60 \\
3C~279 & 1.87 & 1.12 \\
PG~1407+265 & 2.29 & 1.37 \\
PKS~2135-147 & 2.36 & 1.02 \\
\hline
\end{tabular}
\parbox{3.5in}{
\vspace{0.1in}
a. All column densities are in units of $10^{16}\rm\,cm^{-2}$.\\
}
\end{center}
}

Using standard cosmological models and ingredients from galaxy and
large scale structure evolution, numerical simulations predict the
spatial distribution of metals and their ionization structures (see,
e.g., \citealp{cos99}). Such simulations allow us to make 
quantitative predictions
of the absorption features that could be present in the spectra of
background sources (see, e.g.,
\citealp{hgm98,fbc02,che03}). Our non-detection of any significant
absorption lines can, conveniently, put constraints on
cosmological parameters and other physical processes adopted in
the simulations.

Rather than running complicated hydrodynamic simulations with various
cosmological parameters and physical processes to explore the
parameter space that can be constrained by our observation results,
we adopted a simple but effective analytic approach that follows
\citet{per98} and \citet{fan00}. The basic idea is that all
the hot gas is distributed within virialized halos that follow a
Press-Schechter distribution \citep{pre74}. Given a
certain density profile of a virialized halo, the probability that a
random line-of-sight that passes through a halo of mass $M$ with an impact
parameter $b$ can be calculated. Based on assumptions of metal
abundance, we can then calculate the so-called ``X-ray Forest
Distribution Function'' (XFDF), defined as $\partial^2P/(\partial
N_i\partial z)$, the
absorption line number per unit redshift ($z$) per column density
($N_i$) for ion
species $i$. Specifically, XFDF can be analytically calculated as
\begin{equation}
\frac{\partial^2P}{\partial N_i\partial z} = \int_M
\frac{dn}{dM}\frac{d\Sigma}{dN_i}\frac{dl}{dz},
\end{equation}
where $\left(dn/dM\right)$ is the distribution of the comoving
virialized halo,  $\Sigma$ is the cross section of the halo, and $l$
is the path length.

How accurate is this analytic approach? \citet{fbc02} compared both 
numerical and semi-analytic methods (with the above
analytic model and a halo temperature profile fitted from
simulations). For \ion{O}{7} and \ion{O}{8}, it turns out that there
is a large discrepancy in the low column density region ($N_i \sim 10^{12} -
10^{15}\,\rm cm^{-2}$), where most lines are distributed in the
filamentary structures seen in the numerical simulations but which 
cannot be described by Press-Schechter
formalism. At the high column density end ($N_i \gtrsim 10^{15}\,\rm cm^{-2}$),
the semi-analytic method provides a reasonablely good fit to
results from numerical simulations. These high column density lines
distribute typically in virialized halos with higher temperatures and
densities than found in the filaments. Since in our study the minimum 
detectable column densities
are around $10^{16}\,\rm cm^{-2}$, the analytic estimate is appropriate.

Given the XFDF and certain cosmological parameters, we can calculate
the expected absorption line number by combining observations on all
the six targets. The total absorption line number is
then
\begin{equation}
n = \sum_j \int_{N_i(j)}^\infty \int_0^{z_j} 
\frac{\partial^2P}{\partial N_i\partial z} dN_i dz.
\end{equation}
Here $N_i(j)$ is the minimum detectable column density of ion $i$ for
the $j$th target; $z_j$ is its maximum redshift that can probed with
the {\sl Chandra} and {\sl XMM}-Newton instruments; and overall
summation is over all the six targets (index $j$).

The parameters we plan to constrain are the cosmic matter density
$\Omega_m$ and metal abundance $Z$ (in units of the solar abundance 
$Z_\odot$.) Changes in these two
parameters will dramatically change gas and metal content and could
have significantly impact on the detectability of X-ray absorption
lines. We keep all the other parameters at the standard values: we use
the $\Lambda$CDM model with a dark energy density of
$\Omega_{\Lambda}=1-\Omega_m$, the Hubble constant is
$H_0=100h\,\rm km\,s^{-1}Mpc^{-1}$ with $h=0.67$; the baryon
density is $\Omega_b=0.04$; and we assume the gas fraction, or the
ratio of baryonic-to-total mass, is $f_{gas}
= 0.2$.

\vskip0.5cm
\begin{figurehere}
\psfig{file=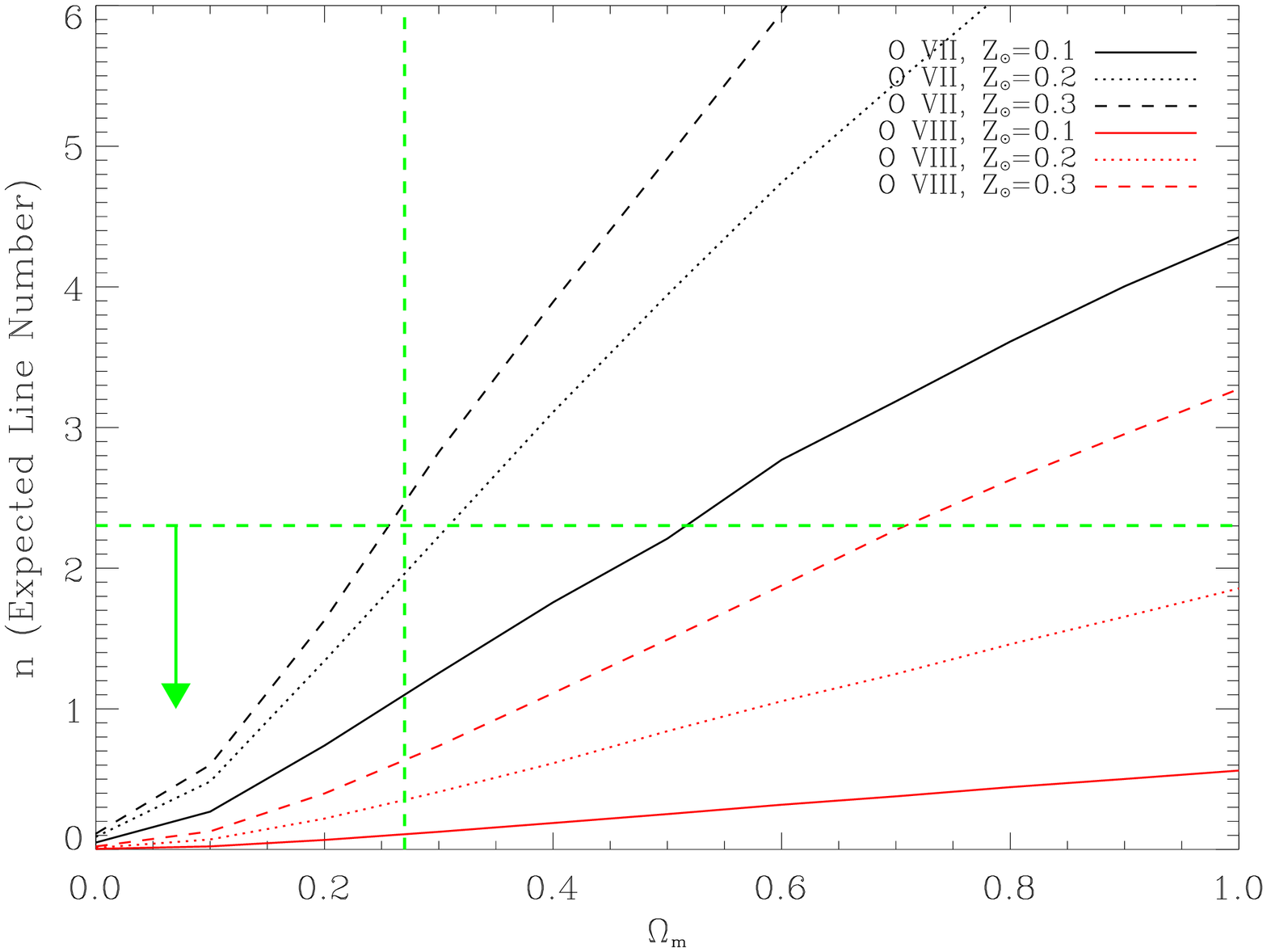,width=0.35\textwidth,height=0.33\textheight,angle=90}
\caption[h]{the expected total \ion{O}{7} (dark
lines) and \ion{O}{8} (red lines)  line numbers with different
$\Omega_m$ and metal abundance $Z$. The horizontal green line shows the
90 \% confidence upper limit from the non detection of absorption line in our
sample. The vertical green line shows the value of $\Omega_m$
measured with {\sl WMAP} \citep{spe03}. The green arrow
indicates allowed region.}
\label{f11}
\end{figurehere}
\vskip0.5cm

Figure~\ref{f11} shows the expected total \ion{O}{7} (black
lines) and \ion{O}{8} (red lines)  line numbers with different
$\Omega_m$ and metal abundance $Z$. Since we detect no absorption
line, the horizontal green line shows the 90\% confidence upper limit
\citep{geh86}. The vertical green line shows the value of $\Omega_m$
measured with {\sl WMAP} \citep{spe03}. The green arrow
indicates allowed region. Clearly, we expect to observe more
absorption lines with higher values of metallicity and $\Omega_m$. The
degeneracy between metallicity and $\Omega_m$ can be broken by
applying the much more accurate value of $\Omega_m$ measured with {\sl
   WMAP} \citep{spe03}: $\Omega_mh^2=0.14$, indicated by the
vertical green line in Figure~\ref{f11}. We cannot constrain
metallicity of the IGM based on the measurement of \ion{O}{8}, i.e.,
the predicted line number is still less than the 90\% upper
limit from our observation even if the metallicity is as high as
0.3$Z_{\odot}$. However, \ion{O}{7} can put much better constraint on
the IGM metallicity: for a {\sl WMAP} value of $\Omega_m$, the metal
abundance should be smaller than $ \sim 0.3 Z_\odot$. That \ion{O}{7} 
is more restrictive is not surprising: in the collisionally ionized 
WHIM gas, simulations showed
that \ion{O}{7} is more abundant than \ion{O}{8} and should produce
more absorption lines (see Figure~13 of \citealp{fbc02}).

\subsection{Local ($z \approx 0$) Absorption}

Recently, there are a number of observations of quasars showing local 
($z \approx 0$)
X-ray absorption lines along several line-of-sight. These
background quasars (see,
\citealp{kas02,nic02,fsc03,ras03,cag04,mck04}) are among the
brightest extragalactic X-ray sources in the sky (some of them  
are used as instrument calibration targets). In particular,
\citet{mck04} conducted a survey of nearby type I AGNs, and found
that about half of their sample exhibit local ($z \approx 0$)
absorption from H- or He-like oxygen. The high column
densities of these X-ray absorbers ($N \sim 5~\times 10^{15} - 2 
\times 10^{16}\rm\,
cm^{-2}$ for \ion{O}{7}) imply the existence of large amounts of hot 
gas. Given the
current instrument resolution with {\sl Chandra} and {\sl XMM}-Newton,
velocity measurements cannot distinguish its location: in the interstellar
medium, in the distant halo, or in the Local Group as the intragroup
medium.

No local, $z \approx 0$, absorption lines are seen in our sample. We
also notice none of these targets exhibit local absorption from
\ion{O}{6} in UV band \citep{sem03} The limits, from Table~4, are
comparable to  or below some of the lines detected along other lines
of sight.  A  systematic study of detections and non-detections and
their  implications will be presented in a forthcoming
publication. Here we  note that, whatever its location, the column
density of local  absorption fluctuates at least by a factor of two
across the sky.  More specifically, for 3C~279 our 3~$\sigma$ limit of
$1.1 \times  10^{16}\,\rm cm^{-2}$ for \ion{O}{7} is well below the
detected value  of $\sim 1.8\times 10^{16}\,\rm cm^{-2}$ for 3C~273
\citep{fsc03}.  These two sources are separated by an angular distance
of $ \sim  10\arcdeg$. This corresponds to a linear scale of absorber
structure  of $\Delta l \lesssim 90 (D/500\rm\,kpc)$ kpc, for a
distance to the absorber D. \footnote{A distance of $500$ kpc is
likely an upper limit. It is unlikely that the 3c~273 absorber has a
distance from us of $\gtrsim 1$ Mpc since it will then be located even
beyond our Local Group, see \citet{fsc03} on discussion of the
geometry of the Local Group and 3C~273 absorber.}. If the absorber is
clumped on this scale, we can obtain a lower limit on the hydrogen
number density of
\begin{equation}
n_H \gtrsim 4\times 10^{-3} \left(\frac{Z_O}{0.1\,Z_{\odot}}\right)^{-1}
\left(\frac{D}{500\,\rm kpc}\right)^{-1}\,\rm cm^{-3}.
\label{eq:nH}
\end{equation} Here we adopt a
metalicity of $Z_O \sim 0.1\,Z_{\odot}$ and assume half the oxygen is 
ionized to \ion{O}{7}.

\section{Summary} \label{sec:sum}

In this paper we present our {\sl Chandra} and {\sl XMM}-Newton
observations of six quasars, with redshifts ranging from $\sim$ 0.1 to
$\sim$ 0.9. Our main conclusions can be summarized as follows.

\begin{enumerate}

\item We obtained the continuum information of these six targets. All
   of them can be fitted quite well with a power law with absorption
   from neutral hydrogen, although some of these targets may require
   additional components to achieve a better fit.

\item Our main purpose is to search for any narrow absorption features
   in the X-ray spectra that were produced by the WHIM gas. After
   applying an optimized local fitting technique to subtract continuum,
   we found a total of 30 narrow features at or above $3\sigma$
   lever. A cross check with different instruments excludes all
   features. Thus we conclude that no
   absorption feature was detected in our observations.

\item Given the size of our survey, we are able to put
   stronger constraints on cosmological parameters, compared with
   previous studies (see, e.g., \citealp{fan00}). We find the metal
   abundance of the IGM must be smaller than $ \sim 0.3 Z_{\odot}$ if
   we adopted a {\sl WMAP} value of $\Omega_m$.

\item We also search for local ($z \sim 0$) absorption lines, such as 
those that
   have been detected along several other lines of sight. No local 
absorption line
   was found.

\item The limit on a ($z \sim 0$) absorber in 3C~279 compared with 
the detection of a strong absorption line in nearby 3C~273 indicates 
that the absorber has an angular scale of $ \sim 10 \arcdeg$.  If the 
3C~273 absorber is in a cloud of this scale we estimate a lower limit 
on the hydrogen number density of $n_H \approx 4\times 10^{-3}\,\rm 
cm^{-3}$.  Implications will be explored further in a forthcoming 
publication.

\end{enumerate}

Though several lines of evidence imply the existence of the WHIM
gas, firm evidence for such ``missing baryons'' still remains an
important challenge in cosmology. Future X-ray missions are very
promising for detecting WHIM gas (see, e.g., \citealp{che03}). Both
{\sl Constellation}-X \footnote{See
   http://constellation.gsfc.nasa.gov/} and {\sl XEUS} \footnote{See 
http://www.rssd.esa.int/index.php?project=XEUS.} can probe highly 
ionized metals
to column densities of as low as $10^{14}$ - $10^{15}\rm\,
cm^{-2}$. Such high resolution and sensitivities will reveal a true
``X-ray Forest'' in the IGM. By comparing observations with numerical
simulations and analytic analysis, we may eventually obtain a complete
theory of structure formation and evolution from the big bang to the
present epoch.

\smallskip
{\it Acknowledgments:} This work was support under NASA contract
8-38249 and {\sl XMM}-Newton GO grant 6891353. TF was supported by the
NASA through {\sl Chandra} Postdoctoral Fellowship Award Number
PF3-40030 issued by the {\sl Chandra} X-ray Observatory Center, which
is operated by the Smithsonian Astrophysical Observatory for and on
behalf of the NASA under contract NAS8-39073. HLM was also supported
under SAO SV1-61010 for the Chandra X-Ray Center.


\begin{thebibliography}{99}

\bibitem[Blustin, Page, \& Branduardi-Raymont(2004)]{blu04}
Blustin, A.~J., Page, M.~J., \& Branduardi-Raymont, G.\ 2004, A\&A, 417, 61

\bibitem[Cagnoni et al.(2004)]{cag04} Cagnoni, I.,
   Nicastro, F., Maraschi, L., Treves, A., \& Tavecchio, F.\ 2004,
   ApJ, 603, 449

\bibitem[Canizares et al.(2005)]{can05} Canizares, C.~R.\ 2005,
PASP, submitted

\bibitem[Cash(1976)]{cas76} Cash, W.\ 1976, A\&A, 52, 307 

\bibitem[Cen \& Ostriker(1999)]{cos99} Cen, R.~\& Ostriker, J.~P.\ 
1999, ApJ, 514, 1

\bibitem[Chen, Weinberg, Katz, \& Dav{\' e}(2003)]{che03}
Chen, X., Weinberg, D.~H., Katz, N., \& Dav{\' e}, R.\ 2003, \apj, 594, 42

\bibitem[Comastri, Fossati, Ghisellini, \&
Molendi(1997)]{cfg97} Comastri, A., Fossati, G., Ghisellini,
G., \& Molendi, S.\ 1997, \apj, 480, 534

\bibitem[Costamante et al.(2001)]{cos01} Costamante, L.~et
al.\ 2001, \aap, 371, 512

\bibitem[Cox(2000)]{cox00} Cox, A.~N.\ 2000, Allen's
astrophysical quantities, 4th ed.~ Publisher: New York: AIP Press;
Springer, 2000.~ Edited by Arthur N.~Cox.~ ISBN: 0387987460

\bibitem[Dav{\' e} et al.(2001)]{dav01} Dav{\' e}, R.,
   et al.\ 2001, \apj, 552, 473

\bibitem[Falomo, Pesce, \& Treves(1993)]{fpt93} Falomo, R.,
Pesce, J.~E., \& Treves, A.\ 1993, \aj, 105, 2031

\bibitem[Fang, Bryan, \& Canizares(2002)]{fbc02} Fang, T.,
Bryan, G.~L., \& Canizares, C.~R.\ 2002, \apj, 564, 604

\bibitem[Fang \& Canizares(2000)]{fan00} Fang, T.~\&
   Canizares, C.~R.\ 2000, \apj, 539, 532

\bibitem[Fang et al.(2002a)]{fan02a} Fang, T., Marshall, H.~L.,  Lee,
J.~C., Davis, D.~S., \& Canizares, C.~R.\ 2002, \apjl, 572, L127

\bibitem[Fang et al.(2002b)]{fan02b} Fang, T., Davis,
   D.~S., Lee, J.~C., Marshall, H.~L., Bryan, G.~L., \& Canizares,
   C.~R.\ 2002, \apj, 565, 86

\bibitem[Fang, Marshall, Bryan, \&
   Canizares(2001)]{fan01} Fang, T., Marshall, H.~L.,
   Bryan, G.~L., \& Canizares, C.~R.\ 2001, \apj, 555, 356

\bibitem[Fang, Sembach, \& Canizares(2003)]{fsc03} Fang, T.,  Sembach,
K.~R., \& Canizares, C.~R.\ 2003, \apjl, 586, L49

\bibitem[Fang et al.(2005)]{fan05} Fang, T., et al.\ 2005, \apj, in
  press 

\bibitem[Ganguly et al.(2001)]{gan01} Ganguly, R., Bond,
N.~A., Charlton, J.~C., Eracleous, M., Brandt, W.~N., \& Churchill, C.~W.\
2001, \apj, 549, 133

\bibitem[Garmire et al.(2003)]{gar03} Garmire, G.~P.,
   Bautz, M.~W., Ford, P.~G., Nousek, J.~A., \& Ricker, G.~R.\ 2003,
   \procspie, 4851,
28

\bibitem[Gehrels(1986)]{geh86} Gehrels, N.\ 1986, \apj, 303,
336

\bibitem[Gower \& Hutchings(1984)]{ghu84} Gower, A.~C.~\&
Hutchings, J.~B.\ 1984, \aj, 89, 1658

\bibitem[Hamann et al.(1997)]{ham97} Hamann, F., Beaver,
E.~A., Cohen, R.~D., Junkkarinen, V., Lyons, R.~W., \& Burbidge, E.~M.\
1997, \apj, 488, 155

\bibitem[Hellsten, Gnedin, \& Miralda-Escud{\'
e}(1998)]{hgm98} Hellsten, U., Gnedin, N.~Y., \&
Miralda-Escud{\' e}, J.\ 1998, \apj, 509, 56

\bibitem[Houck \& Denicola(2000)]{hde00} Houck, J.~C.~\& Denicola,
L.~A.\ 2000, ASP Conf.~Ser.~216: Astronomical Data Analysis Software
and Systems IX, 9, 591

\bibitem[Kaspi et al.(2002)]{kas02} Kaspi, S., et al.\ 2002,
\apj, 574, 643

\bibitem[Lanzetta, Wolfe, \& Turnshek(1995)]{lwt95} Lanzetta,
K.~M., Wolfe, A.~M., \& Turnshek, D.~A.\ 1995, \apj, 440, 435

\bibitem[Madejski et al.(1992)]{mad92} Madejski, G.~et al.\
1992, Frontiers Science Series, 583

\bibitem[Mathur, Weinberg, \& Chen(2002)]{mwc02} Mathur, S., Weinberg,
D., \& Chen, X.\ 2002, \apj, 582, 82

\bibitem[Marshall, Cheung, Canizares, \& Fang(2003)]{mar03}
Marshall, H.~L., Cheung, T., Canizares, C.~R., \& Fang, T.\ 2003, American
Astronomical Society Meeting, 202

\bibitem[McKernan et al.(2003)]{mck03} McKernan, B.,
   Yaqoob, T., Mushotzky, R., George, I.~M., \& Turner, T.~J.\ 2003,
   \apjl, 598, L83

\bibitem[McKernan et al.(2004)]{mck04} McKernan, B., Yaqoob, 
T., \& Reynolds, C.~S.\ 2004, \apj, 617, 232

\bibitem[Miley \& Hartsuijker(1978)]{mha78} Miley, G.~K.~\&
Hartsuijker, A.~P.\ 1978, \aaps, 34, 129

\bibitem[Nicastro et al.(2002)]{nic02} Nicastro, F.~et al.\ 2002,
\apj, 573, 157

\bibitem[Nicastro et al.(2005)]{nic05} Nicastro, F.,
et al.\  2005, \nat, 433, 495 

\bibitem[Perna \& Loeb(1998)]{per98} Perna, R.~\& Loeb,
   A.\ 1998, \apjl, 503, L135

\bibitem[Polomski et al.(1997)]{pol97} Polomski, E., Vennes,
S., Thorstensen, J.~R., Mathioudakis, M., \& Falco, E.~E.\ 1997, \apj, 486,
179

\bibitem[Press \& Schechter(1974)]{pre74} Press,
   W.~H.~\& Schechter, P.\ 1974, \apj, 187, 425

\bibitem[Rachen, Mannheim, \& Biermann(1996)]{rmb96} Rachen,
J.~P., Mannheim, K., \& Biermann, P.~L.\ 1996, \aap, 310, 371

\bibitem[Rasmussen, Kahn, \& Paerels(2003)]{ras03} Rasmussen,
A., Kahn, S.~M., \& Paerels, F.\ 2003, ASSL Vol.~281: The IGM/Galaxy
Connection.~The Distribution of Baryons at z=0, 109

\bibitem[Reeves \& Turner(2000)]{rtu00} Reeves, J.~N.~\&
Turner, M.~J.~L.\ 2000, \mnras, 316, 234

\bibitem[Remillard et al.(1989)]{rem89} Remillard, R.~A.,
Tuohy, I.~R., Brissenden, R.~J.~V., Buckley, D.~A.~H., Schwartz, D.~A.,
Feigelson, E.~D., \& Tapia, S.\ 1989, \apj, 345, 140

\bibitem[Sambruna(1997)]{sam97} Sambruna, R.~M.\ 1997, \apj,
487, 536

\bibitem[Sambruna et al.(1997)]{sam97a} Sambruna, R.~M.,
George, I.~M., Madejski, G., Urry, C.~M., Turner, T.~J., Weaver, K.~A.,
Maraschi, L., \& Treves, A.\ 1997, \apj, 483, 774

\bibitem[Savage, Tripp, \& Lu(1998)]{stl98} Savage, B.~D.,  Tripp,
T.~M., \& Lu, L.\ 1998, \aj, 115, 436

\bibitem[Sembach et al.(2003)]{sem03} Sembach, K.~R., et al.\ 
2003, \apjs, 146, 165 

\bibitem[Simcoe, Sargent, \& Rauch(2002)]{ssr02} Simcoe,  R.~A.,
Sargent, W.~L.~W., \& Rauch, M.\ 2002, \apj, 578, 737

\bibitem[Singh, Rao, \& Vahia(1991)]{srv91} Singh, K.~P.,
Rao, A.~R., \& Vahia, M.~N.\ 1991, \aap, 243, 67

\bibitem[Spergel et al.(2003)]{spe03} Spergel, D.~N., et al.\
2003, \apjs, 148, 175

\bibitem[Spitzer(1978)]{spi78} Spitzer, L.\ 1978, New
   York Wiley-Interscience, 1978, p.~333

\bibitem[Tripp \& Savage(2000)]{tsa00} Tripp, T.~M.~\&  Savage, B.~D.\
2000, \apj, 542, 42

\bibitem[Tripp, Savage, \& Jenkins(2000)]{tsj00} Tripp,  T.~M.,
Savage, B.~D., \& Jenkins, E.~B.\ 2000, \apjl, 534, L1

\bibitem[Ueda et al.(2001)]{ued01} Ueda, Y., Ishisaki, Y.,
Takahashi, T., Makishima, K., \& Ohashi, T.\ 2001, \apjs, 133, 1

\bibitem[Verner, Verner, \& Ferland(1996)]{ver96}
   Verner, D.~A., Verner, E.~M., \& Ferland, G.~J.\ 1996, Atomic Data
   and Nuclear Data Tables, 64, 1

\bibitem[Viel et al.(2003)]{vie03} Viel, M., Branchini,
   E., Cen, R., Matarrese, S., Mazzotta, P., \& Ostriker, J.~P.\
   2003, \mnras, 341, 792

\bibitem[Whitney et al.(1971)]{whi71} Whitney, A.~R.~et al.\
1971, \baas, 3, 465

\bibitem[Wilkes \& Elvis(1987)]{wel87} Wilkes, B.~J.~\&
Elvis, M.\ 1987, \apj, 323, 243

\bibitem[Williams et al.(2005)]{wil05} Williams, R. et al.\ 2005,
  \apj, submitted (astro-ph/0504558)

\bibitem[Wolter et al.(1998)]{wol98} Wolter, A.~et al.\ 1998,
\aap, 335, 899

\bibitem[Wood et al.(1984)]{woo84} Wood, K.~S.~et al.\ 1984,
\apjs, 56, 507

\bibitem[Yoshikawa et al.(2003)]{yos03} Yoshikawa, K., 
Yamasaki, N.~Y., Suto, Y., Ohashi, T., Mitsuda, K., Tawara, Y., \& 
Furuzawa, A.\ 2003, \pasj, 55, 879

\end{thebibliography}
\end{document}